\documentclass[doublecol,figures]{epl2}

\usepackage{amssymb,amsmath}
\usepackage{epsf,colordvi,graphicx,color,amsbsy}
\usepackage[T1]{fontenc}
\usepackage{ulem}

\title{Equation of state of a granular gas homogeneously driven by particle rotations}
\author{E. Falcon\thanks{Corresponding author: \email{eric.falcon@univ-paris-diderot.fr}}, J.-C. Bacri \and C. Laroche}
\shortauthor{E. Falcon, J.-C. Bacri \and C. Laroche}
\institute{Univ Paris Diderot, Sorbonne Paris Cit\'e, MSC, UMR 7057 CNRS, F-75 013 Paris, France, EU}
\abstract{
We report an experimental study of a dilute ``gas'' of magnetic particles subjected to a vertical alternating magnetic field in a 3D container. Due to the torque exerted by the field on the magnetic moment of each particle, a spatially homogeneous and chaotic forcing is reached where only rotational motions are driven. This forcing differs significantly from boundary-driven systems used in most previous experimental studies on non equilibrium dissipative granular gases. Here, no cluster formation occurs, and the equation of state displays strong analogy with the usual gas one apart from a geometric factor. Collision statistics is also measured and shows an exponential tail for the particle velocity distribution. Most of these observations are well explained by a simple model which uncovers out-of-equilibrium systems undergoing uniform ``heating''.}     


\begin{document}
\maketitle
  
\section{Introduction}
 Granular gases display striking properties compared to molecular gases: cluster formation at high enough density \cite{Gollub97,Falcon99,Falcon99b}, anomalous scaling of pressure \cite{Falcon99,Falcon99b} and collision frequency \cite{Falcon06}, non-Gaussian distribution of particle velocity \cite{Rouyer00}. These differences are mainly ascribed to dissipation occurring during inelastic collisions between particles. A continuous input of energy is thus required to reach a non equilibrium steady state for a granular gas. This is usually performed experimentally by vibrating a container wall or the whole container. For such vibration-fluidized systems, the role of the boundary condition affects the shape of the particle velocity distribution \cite{Rouyer00}, as well as the extent of energy nonequipartition \cite{Wang08}. A spatially homogeneous forcing, driving each particles stochastically, is thus needed to explore the validity domain of granular gas theories. However, it is hardly reachable in experiments \cite{Cafiero02}. Here, we study experimentally the equation of state and the collision statistics of a spatially homogeneous driven granular gas in a 3D container. Magnetic particles subjected to a magnetic field oscillating in time are used to homogeneously and stochastically drive the system by injecting rotational energy into each particle. Rotational motion is transferred to translational motion by the collisions with boundaries or other particles. To our knowledge, this type of forcing has been only used to study magnetic particles in pattern formation, in suspensions on liquid surface \cite{Snezhko05}, or to measure their velocity distribution in 2D cells \cite{Schmick08}.  Beyond direct interest in out-of-equilibrium statistical physics, granular medium physics, and geophysics (such as dust clouds or planetary rings \cite{Goldreich78}), our study provides an insight into applied problems such as magnetic hyperthermia for medical therapy  \cite{Lesfilles} or electromagnetic grinders in steel mills \cite{Lupanov07}, where particle dynamics are controlled by an alternating magnetic field. 
 
 \begin{figure}[ht!]
    \centering
        \includegraphics[height=75mm]{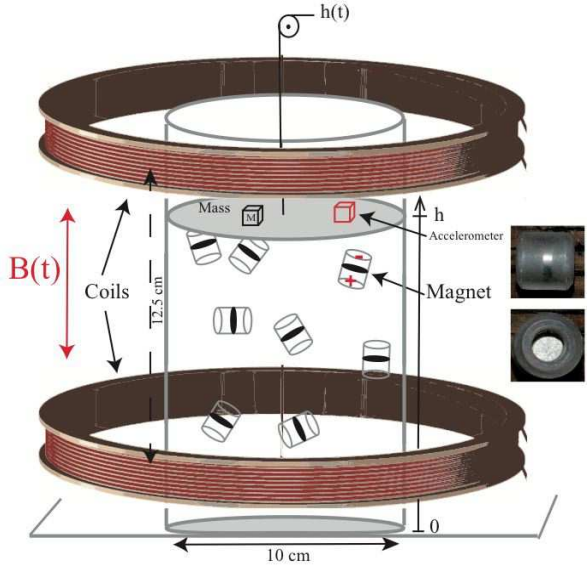}
    \caption{(Color online) Experimental set up. Right insets show pictures of a magnetic particle (1 cm scale).}
    \label{fig01}
\end{figure}
\section{Experimental setup} The experimental setup is shown in Fig.\ \ref{fig01}. A cylindrical glass container, 10 cm in diameter and 14 cm in height, is filled with $N$ magnetic particles, with $2\leq N \leq 60$ corresponding to less than 1 layer of particles at rest. 
Magnetic particles are constituted of a disc permanent magnet in Neodymium (NdFeB, N52, 0.5 cm in diameter and 0.2 cm in thickness) encased and axially aligned in a home made plexiglas cylinder ($d$=1 cm in outer diameter, 0.25 cm in thickness, and $L=1$ cm long) -- see pictures in Fig.\ \ref{fig01}. The aim of this casing is to strongly reduce by factor $38$ the dipole-dipole interaction between two particles compared to the case of magnets without casing. %This configuration enables a reduction of a factor $38$ of the dipole-dipole interaction between two particles. 
The magnetic induction of this dipolar particle, $\mu_0 \mathcal{M}=250$~G, was measured by a Hall probe at the top of the cylinder, where $\mathcal{M}$ is the magnetization of the particle, and $\mu_0=4\pi 10^{-7}$ H/m. Its magnetic moment is $\mathsf{m} \equiv \mathcal{M}V_p$ with $V_p=\pi d^2L/4=0.78$ cm$^3$ the volume of a particle.  The container is aligned between two coaxial coils, 18 cm (40 cm) in inner (outer) diameter, 12.5 cm apart, as shown in Fig.\ \ref{fig01}. A 50 Hz alternating current is supplied to the coils in series by a variable autotransformer (Variac 260V/20 A). A vertical alternating magnetic induction $B$ is thus generated in the range $0\leq B\leq 225$~G with a frequency $f=50$ Hz. The Helmholtz configuration of the coils ensures a spatially homogeneous $B$ within the container volume with a 3\% accuracy. The motion of particles are visualized with a fast camera (Photron Fastcam SA1.1) at 250 fps or 500 fps. An accelerometer attached to the lid records the particle collisions with the lid for 500 s to extract the collision frequency and the impact amplitude on the lid. The sampling frequency was fixed at $100$ kHz to resolve collisions ($\sim 60$ $\mu$s). We focus here on the dilute regime with volume fractions of $0.2\% \leq \Phi \leq 8\%$, with $\Phi=NV_p/V$ and $V$ the volume of the container. %The particle mean free path is $l\equiv d/\Phi \gtrsim 0.1$ m, and the Knudsen number $Kn\equiv l/h \gtrsim 1$.

\section{Forcing mechanism}
Assume that $\theta$ is the angle between the vertical field $B(t)=B\sin{\omega t}$ and that $\mathsf{m}$ is the magnetic moment of a particle. A torque $\mathsf{m}\times B$ is thus exerted by the field on the magnetic moment of each particle. The angular momentum theorem reads $Id^2\theta(t)/dt^2=\mathsf{m}B\sin{\omega t}\sin{\theta}$, where $I=m(3d^2/4+L^2)/12=0.14$ g\ cm$^2$ is the moment of inertia of the particle, and $m=1$ g, the particle mass. This equation is known to display periodic motions, period doubling, and chaotic motions \cite{Croquette81}. The ratio between the magnetic dipolar energy, $E_d=\mathsf{m}B$, and the rotation energy at the field frequency, $E_{rot}=I\omega^2/2$ controls the stochasticity degree. The synchronization between the angular frequency of the particles and the magnetic field one, $\omega$, is predicted to occur when $E_{d} \ll E_{rot}$, that is $B \ll  I\omega^2/(2\mathsf{m})=493$ G. When this condition is violated, as it is for our magnetic field range, chaotic rotational motion occurs \cite{Croquette81} as shown in the movie N1.m4v for a single particle. The external magnetic field thus generates a chaotic rotational driving of each particle. A spatially homogeneous forcing is thus obtained where only the rotational degrees of freedom of each particle are stochastically driven in time.

\section{Gas-like regime}
\begin{figure}[t!]
 \centering
     \begin{tabular}{cc}
    \includegraphics[width=40mm]{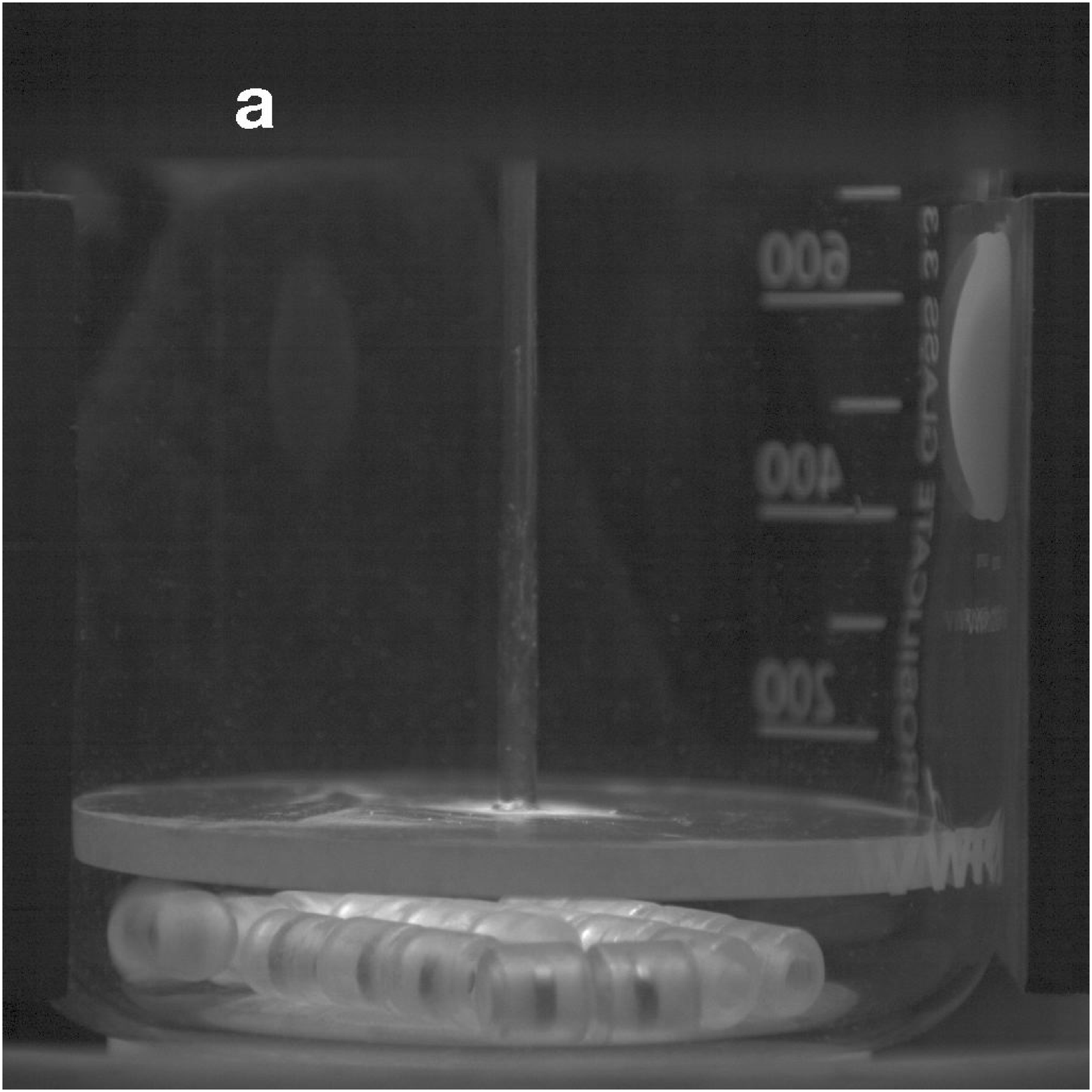} \includegraphics[width=40mm]{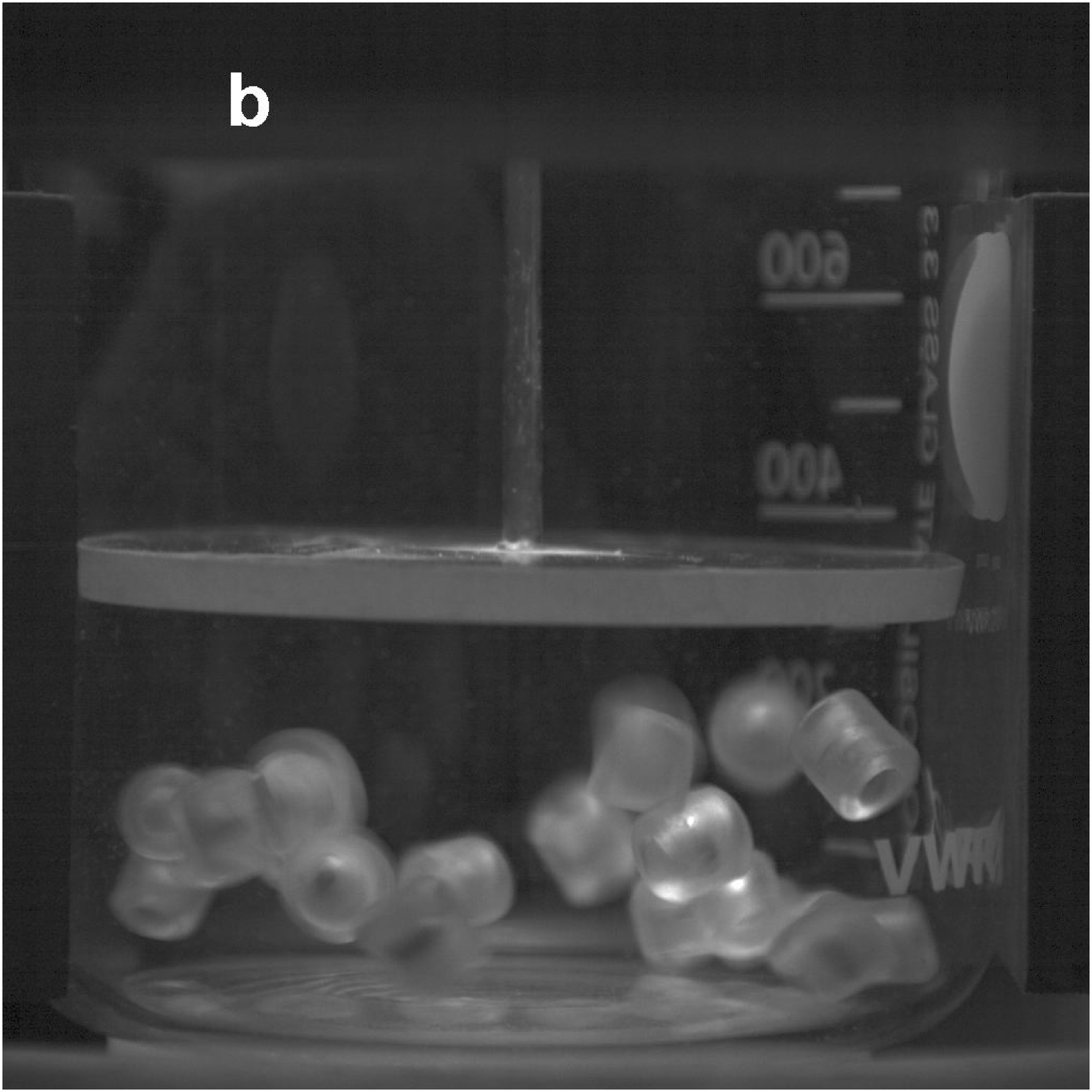}\\
  \includegraphics[width=40mm]{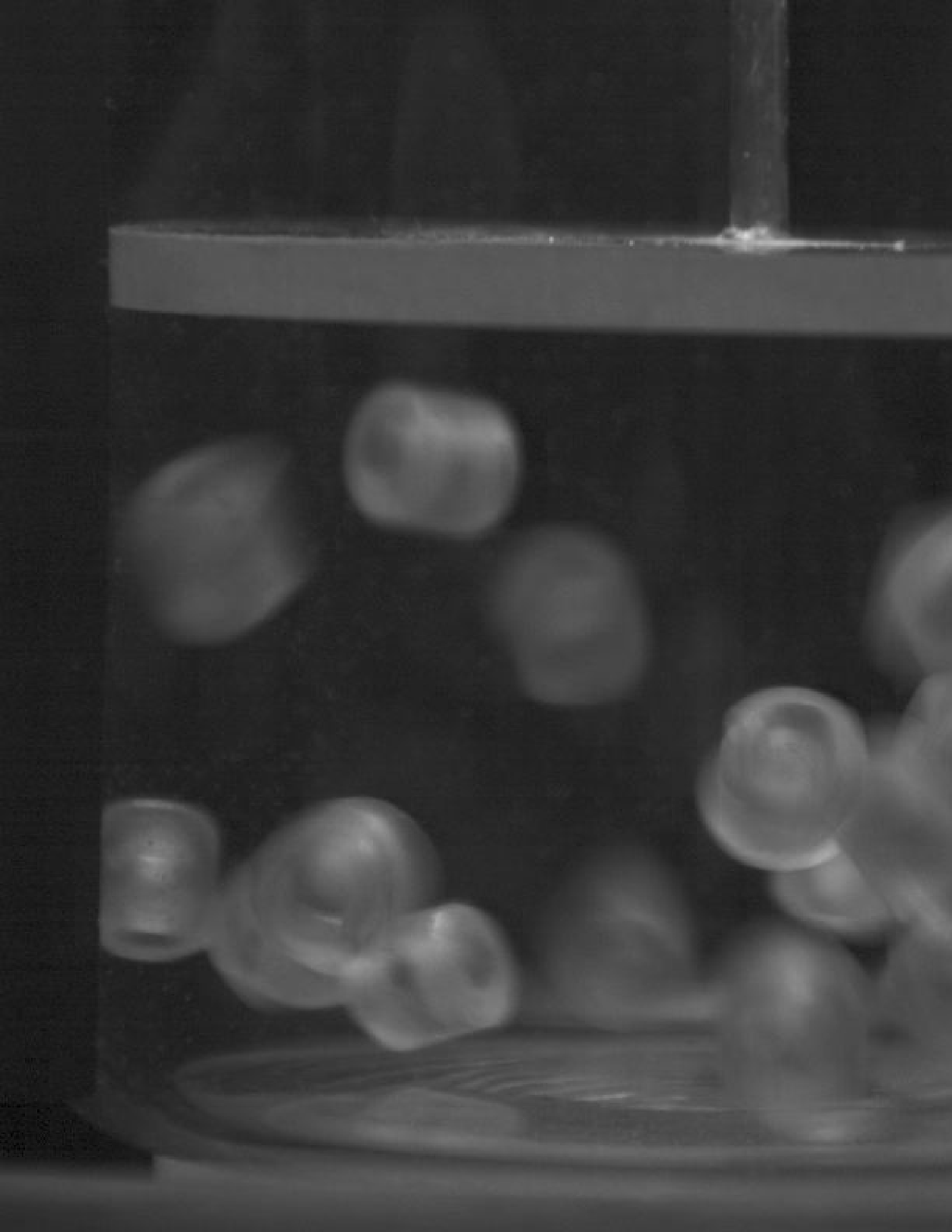} \includegraphics[width=40mm]{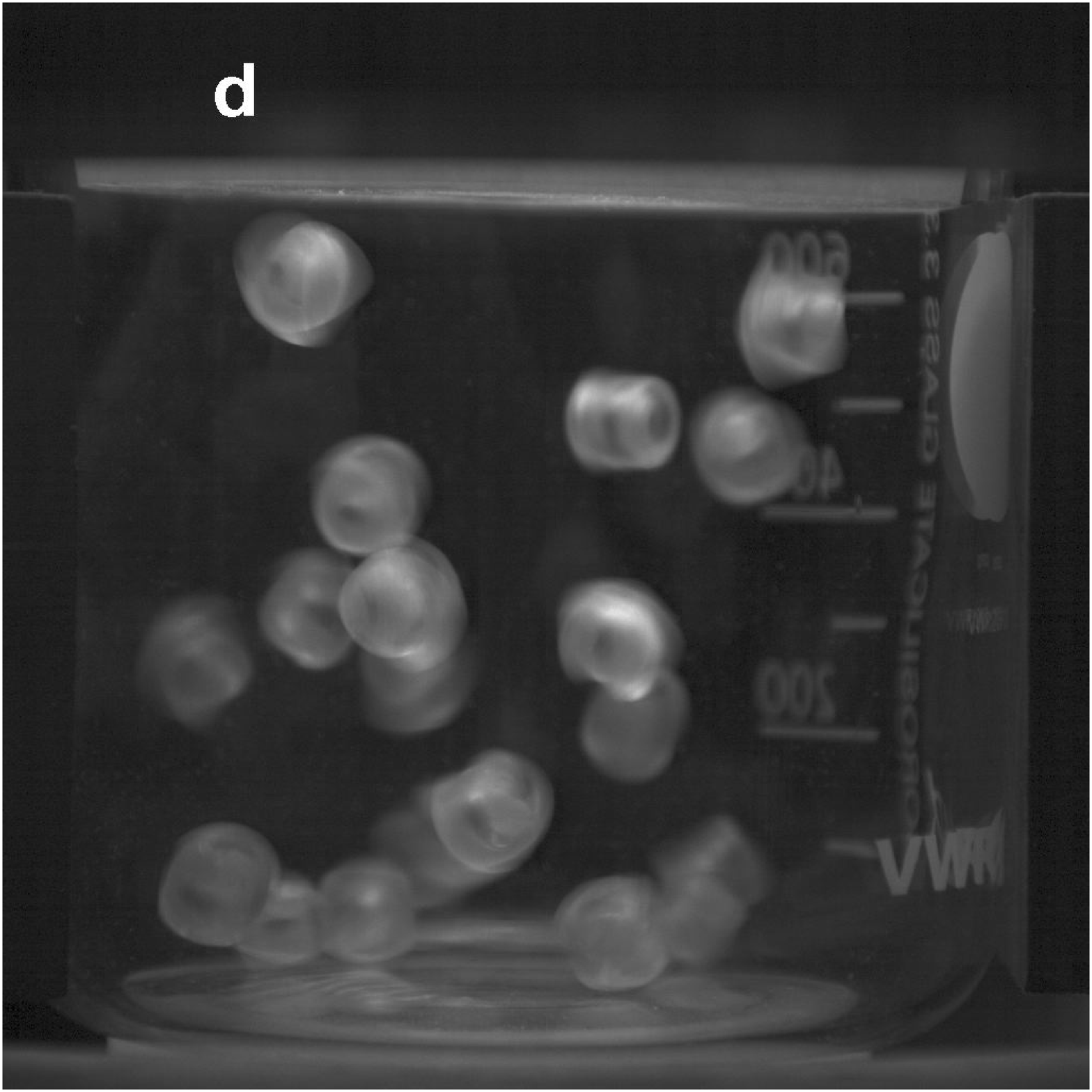}
\end{tabular}
    \caption{(Color online) Snapshots of magnetic granular gas. $N=20$. (a) Initial conditions: $B=0$, a plexiglas lid is laying on the particles. When B is increased from (b) to (d), a gas-like regime develops and the lid is pushed up by the collisions of magnetic particles on it. For the full time evolution, see movie N20onset.m4v}
    \label{fig02}
\end{figure}

$N$ particles are placed at the bottom of the container, their axes lying on the horizontal plane, normal to $B$ (see Fig. \ \ref{fig02}a). A plexiglas lid lays on the particles, and its mass is balanced by a counterweight. When $B$ is increased, a transition occurs at a critical $B_c$: particles begin to jump lifting up the lid. We found that $B_c=75\pm 5$ G regardless of $N$. When $B$ is further increased a stationary gas-like regime is observed with particles rotating and translating erratically -- see Fig.\ \ref{fig02}b-d and movie N20onset.m4v. We observe that the axis of rotation of most particles is normal to the particle axis so their magnetic moments align with the vertical oscillating magnetic field. The frequency and direction of the particle rotation are erratic, showing unpredictably reversals. Their angular frequencies are thus not synchronized with the forcing frequency, $\omega=2\pi f$. The stationary gas-like regime at fixed $B$ is illustrated for $N=10$ and $N=20$ in the movies N10.m4v and N20.m4v (slowed down 100 times and 12.5 times, respectively).

\section{Method}
\begin{figure}[t!]
 \centering
        \includegraphics[height=65mm]{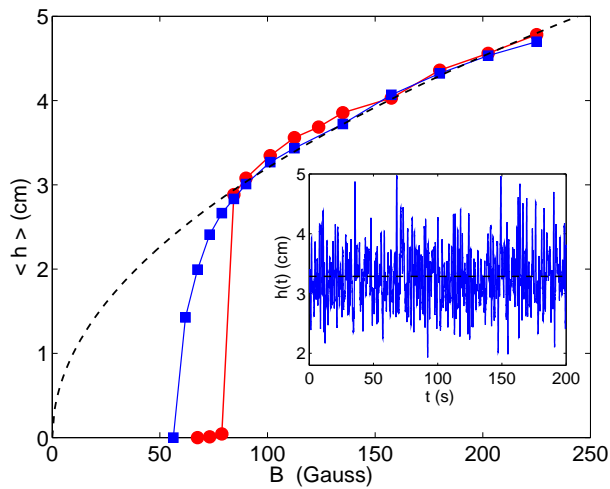} 
    \caption{(Color online) Hysteretic evolution of $\langle h \rangle$ for increasing ($\bullet$) and decreasing ($\blacksquare$) magnetic field $B$. $N=10$, $M=4.7$ g. The dashed line is $\langle h \rangle \sim B^{1/2}$. Inset: Typical temporal evolution of $h(t)$ for $B=101$ G, $N=10$ and $M=4.7$ g. The dashed line is $\langle h \rangle$=3.3 cm.}
    \label{fig03}
\end{figure}

Measurements are performed as follows. A mass $M$ is added on the lid ($0.82 \leq M \leq 10$ g with a 0.82 g step). The lid then stabilizes due to the collisions of particles at a height that depends on $B$ (constant-pressure experiment). The height reached by the lid $h(t)$ exhibits fluctuations in time around a mean height $\langle h \rangle$ as shown in the inset of Fig.\ \ref{fig03}. $h(t)$ is measured by an angular position transducer (12.3 mm/V sensitivity) at a 200 Hz sampling frequency during 200 s. The sensor output voltage is linear with the angle, and the height $h$.  Note that the results reported below are unaffected when performing constant-volume experiments (the lid height is kept constant by adding a mass on the lid that depends on $B$). 

\section{Fluidization onset}
The mean height reached by the lid $\langle h \rangle$ is shown in Fig.\ \ref{fig03} as a function of $B$ for fixed $N$ and $M$. For increasing $B$, a steep jump occurs at the onset $B_c$, whereas a smoother behavior is observed for decreasing $B$. The onset of the particle fluidization is hysteretic, occurring at $B^{i}_c$ for increasing $B$, and at $B^{d}_c<B^{i}_c$ for decreasing $B$. One finds $B^{d}_c=56\pm 1$ G and $B^{i}_c=75\pm 5$ G regardless of $N$. The thresholds come from the balance between the particle magnetic energy, $E_{m}$, and its gravitational energy, $E_g=mgd$, required to lift the particle over one diameter $d$ ($g$ is the acceleration of gravity). When $B$ is decreased, $E_{m}$ corresponds to the particle dipolar energy, $E_d=\mathsf{m}B$, and one finds  $B^{d}_c=mgd/\mathsf{m} \simeq 63$ G. When $B$ is increased, particles are initially in contact, and $E_m$ is the sum of the dipole-dipole interaction energy of two particles in contact, $E_{dd}=\mu_0\mathsf{m}^2/(12V_p)$ \cite{Rosensweig}, and the dipolar energy of a single particle $E_d$. By balancing  $E_{dd} - E_{d}$ with $E_g$, one has $B^{i}_c-B^{d}_c=\mu_0\mathcal{M}/12\simeq 21$ G which matches the experimental value. The hysteresis is thus due to the additional dipole-dipole interaction needed to separate two particles initially in contact. The ratio of the dipolar-dipolar interaction energy, $E_{dd}$, to the dipolar one, $E_d$, reads $E_{dd} / E_{d}=\mu_0 \mathcal{M}/(12B)$. Thus, $E_{dd} \ll E_{d}$ for $B \gg B_c$, whereas $E_{dd} \simeq E_d/3 \simeq E_g/3$ at the onset of fluidization ($B= B^{d}_c$). Thus, the role of dipole-dipole interactions is only limited to the vicinity of the hysteresis.
\begin{figure}[t!]
 \centering
        \includegraphics[height=65mm]{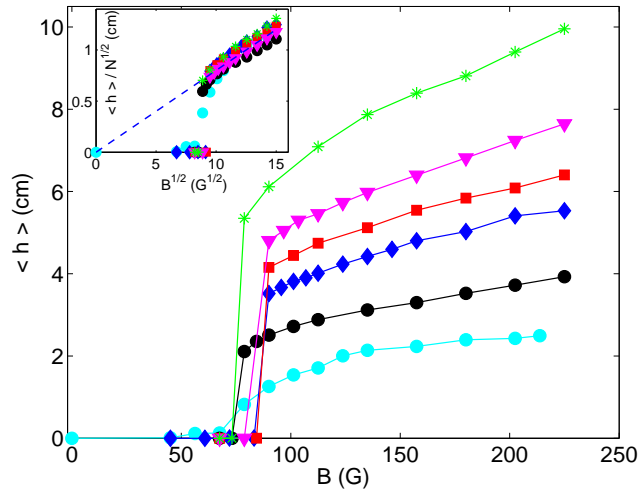}
    \caption{(Color online) $\langle h \rangle$ vs. increasing $B$ for different particle numbers $N=4, 10, 15, 20, 30$ and 40 (from bottom to top). $M=6.9$ g. Inset: best rescaling $\langle h \rangle / N^{1/2}$ vs. $B^{1/2}$. The dashed line has a slope of 0.08  cm/G$^{1/2}$.
    }
    \label{fig04}
\end{figure}
 \begin{figure}[t!]
 \centering
        \includegraphics[height=65mm]{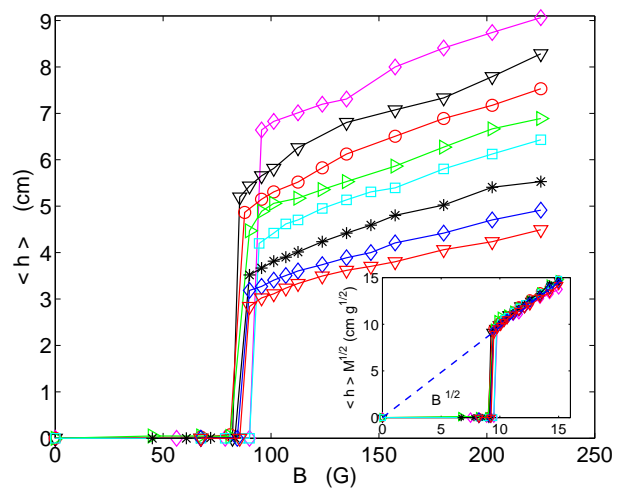}
    \caption{(Color online) $\langle h \rangle$ vs. increasing $B$ for different masses added $M=2.3$, 3.1, 3.9, 4.7, 5.3, 6.9, 8.6 and 10 g (from top to bottom). $N=15$. Inset: best rescaling $\langle h \rangle / M^{1/2}$ vs. $B^{1/2}$. The dashed line has a slope of 0.97  cm g$^{1/2}/$G$^{1/2}$.
    }
    \label{fig05} 
    \end{figure}
    
\section{Equation of state}
Here, we will investigate an empirical equation of state of our system where dipole-dipole interactions are negligible, that is for $B \gg B_c$. Far from the onset, the height reached by the lid is found to scale as $\langle h \rangle \sim B^{x}$ with $x=0.45\pm 0.05$  (see Fig.\ \ref{fig03}). It means that the gaseous regime expands more and more when $B$ increases. Note that a power-law scaling with the onset distance, $\langle h \rangle \sim (B-B_c)^{0.3}$ can be also fitted for decreasing $B$. For fixed $M$, $\langle h \rangle$ is shown in Fig.\ \ref{fig04} as a function of $B$ for different particle numbers $N$. The larger $N$ is, the higher is the height reached by the lid for a fixed $B$. The best rescaling is displayed in the inset of Fig.\ \ref{fig04}, and shows that $\langle h \rangle / N^{1/2} \sim B^{1/2}$. For fixed $N$, $\langle h \rangle$ is shown in Fig.\ \ref{fig05} as a function of $B$ for different added mass $M$ on the lid. The larger $M$ is, the smaller is the height reached by the lid for a fixed $B$. The best rescaling is displayed in the inset of Fig.\ \ref{fig05}, and shows that $\langle h \rangle M^{1/2} \sim B^{1/2}$. To sum up, one finds an experimental state equation for the magnetic granular gas 
\begin{equation}
M \langle h \rangle^2 =k N  B\ {\rm ,} \label{exp}
\end{equation}
where $k=0.05$ g cm$^2/$G is a constant.  
    
\section{Model}
Due to the stochastic forcing, a fraction of the magnetic energy is continuously injected into rotational energy of each particle. Since the out-of-equilibrium system is in a stationary state, the injected energy should be dissipated in average by collisions. A constant exchange of energy occurs during collisions between rotational and translational degrees of freedom as shown in numerical simulations \cite{Brilliantov07}.
%Although the dissipated energy involved rotational and translational parts, both can be assumed to be proportional  \cite{Brilliantov07}. 
Thus, the balance between magnetic energy and translational kinetic energy dissipated during collisions leads to $v^2 \sim \mathsf{m}B$. Accordingly, the typical particle velocity, scales as  
\begin{equation}
v(B) \sim  \sqrt{B}\ {\rm .} \label{vexp}
\end{equation}
More precisely, if we assume simple collision rules (i.e. with no rotation) for the sake of simplicity, the energy loss by a particle of mass $m$ during a collision with the lid of mass $M$ is $\frac{mv^2}{2}(1-\epsilon^2)\frac{M}{m+M}$, where $\epsilon$ is the particle-boundary restitution coefficient. The energy balance finally leads to $v\sim \sqrt{\mathsf{m}B\frac{m+M}{mM(1-\epsilon^2)}}$.

Let us now model the fact that the lid motion under gravity is stabilized at an altitude $h$ due to particle collisions. One thus balances $\tau_l$, the time of flight under gravity of the lid subjected to particle collisions, and $\tau$, the particle time of flight between 2 collisions {\it with the lid} at the height $h$. One has $\tau_l=v_l/g$ with $v_l$ the lid velocity, and $\tau=2h/v$ for $N=1$. For $N$ particles, $\tau$ is given by the experimental results of the next section
\begin{equation}
\tau \sim \frac{h^2}{LNv(B)}\ {\rm ,} \label{tauexp}
\end{equation} 
where $L$ has the dimension of a length, and is experimentally found to be independent of $N$ and $h$ (see below). $L$ is thus the particle size.

%the collision frequency is given, as for an ideal gas, by $1/\tau= v/l$ with $l$ the mean free path \cite{Reif}. Since our Knudsen number is $Kn \equiv l/h \gtrsim 1$, collisions with the walls overwhelm particle-particle ones, and geometrical factors involving the shape of the container has to be considered to compute the collision frequency \cite{Reif}. The easiest way is to divide $\tau=l/v \sim h/v$ by the volume fraction $\Phi=NV_p/V$, where $V=Sh$, and $S$ is the container area. Then, one has
%\begin{equation}
%\tau \sim \frac{h^2S}{Nv(B)V_p}\ {\rm .} \label{tauexp}
%\end{equation} 
Balancing $\tau_l$ with $\tau$ then leads to $h^2 \sim Nvv_lL/g$. The lid velocity is $v_l=v(1+\epsilon)\frac{m}{m+M}$ from simple inelastic collision rules. Thus, using the expressions for $h^2$, $v_l$ and $v$, the theoretical state equation reads
\begin{equation}
Mgh^2 \sim NB\revision{L}\ {\rm ,} \label{EOStheo}
\end{equation}
which is in good agreement with the experimental one of Eq.\ (\ref{exp}). For a more accurate description, complex inelastic collision rules should be included \cite{Huthmann97,Brilliantov07} since linear and angular particle velocities are coupled.
    
     \begin{figure}[t!]
 \centering
        \includegraphics[height=70mm]{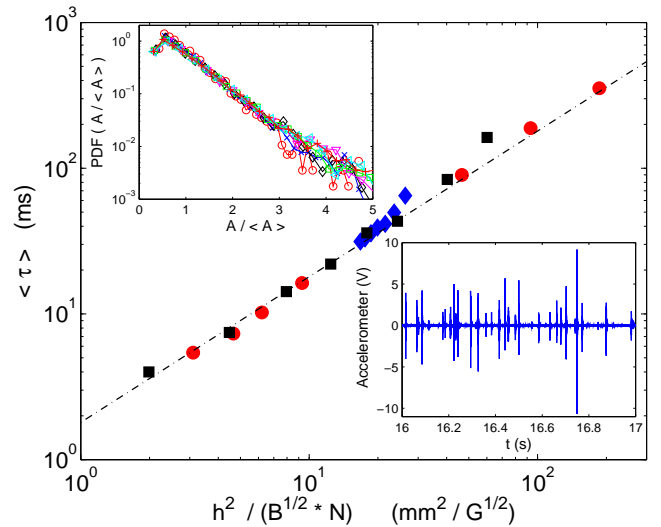} 
    \caption{(Color online) Bottom inset: Temporal signal of the accelerometer showing 35 collisions in 1 s ($N=10$, $h= 5$ cm, $B=157$ G). Main: $\overline{\tau}$ vs. $h^2/(N\sqrt{B})$ for: ($\bullet$) $1\leq N\leq 60$ ($B=180$ G, $h= 5$ cm), ($\blacklozenge$) $90\leq B\leq 222$ G ($N=10$, $h= 5$ cm), and ($\blacksquare$) $2\leq h\leq 11$ cm ($N=15$, $B=180$ G). The dot-dashed line has a unit slope.  Top inset: PDF($A/\overline{A}$) for  $90\leq B\leq 222$ G ($N=15$, $h=5$ cm). }  \label{fig06} 
    \end{figure}
    
\section{Collision statistics}
\begin{figure}[t!]
\centering
\includegraphics[height=70mm]{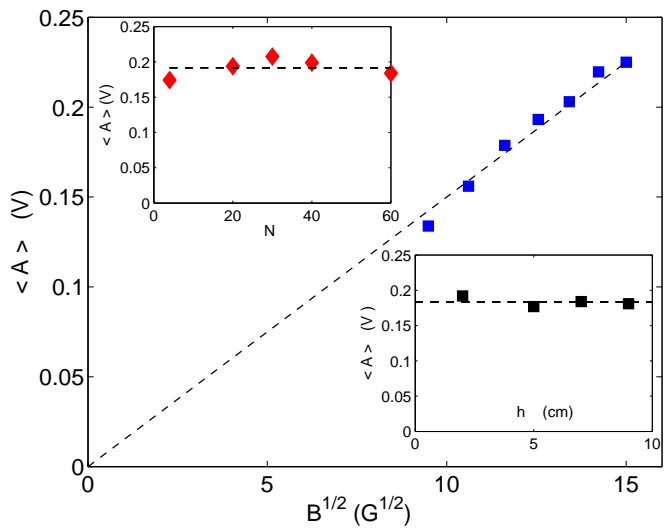} 
\caption{(Color online) Experimental scaling of the mean particle velocity near the wall obtained from the accelerometer measurements. Main: $\overline{A}$ vs. $\sqrt{B}$ for $N=15$, and $h= 6$ cm. Top inset: $\overline{A}$ vs. $N$ for $B=180$ G, and $h=6$ cm. Bottom inset: $\overline{A}$ vs. $h$ for $B=180$ G, and $N=15$.}  \label{fig07} 
\end{figure}

Additional experiments have been performed with the lid fixed to a height $h$. Using an accelerometer attached to the lid, particle collisions with the lid are recorded for $\mathcal{T}=500$ s. A typical acceleration time series is shown in the bottom inset of Fig.\ \ref{fig06}. Each peak corresponds to the acceleration undergoes by a particle during its collision on the lid. The acceleration peak amplitude, $A$, and the time lag, $\tau$, between two successive collisions on the lid are randomly distributed. A thresholding technique is applied to the signal to detect the collisions \cite{Falcon06}. Figure \ref{fig06} shows that the mean time lag scales as $\overline{\tau} =\kappa h^2/\left(NB^{1/2}\right)$ with $\kappa=0.18$ s\ G$^{1/2}/$cm$^2$  over 2 decades when varying one single parameter $h$, $B$ or $N$ while keeping the other two fixed. An experimental verification of Eq.\ (\ref{vexp}) is as follows. The mean amplitude of acceleration peaks is experimentally found to scale as  $\overline{A} \sim h^0N^0B^{1/2}$ as shown in Fig.\ \ref{fig07}. For an impulse response of the accelerometer to a single collision, the product of the acceleration peak amplitude, $A$, times the duration of the collision, $\delta t$, is equal to the magnitude of the particle velocity $v$, and thus $v=A\delta t$ ($\delta t \simeq 60$ $\mu$s is roughly constant). Hence, one has $\overline{v} \sim B^{1/2}$ in agreement with Eq.\ (\ref{vexp}). The translational granular temperature near the wall thus scales as $T_w \sim h^0N^0B^{1}$. 

The number of lid-particle collisions is $N_{coll}=\mathcal{T}/\overline{\tau}$. Typically, $1.5\ 10^3\leq N_{coll} \leq 9\ 10^4$ for $1\leq N \leq 60$ ($\mathcal{T}= 500$ s, $h=5$ cm and $B=180$ G). Although the number of the particle-particle collisions is not measured, it should be much less than the particle-wall ones. Indeed, an estimation of the Knudsen number leads to $Kn \equiv l/h \gtrsim 1$ with $l\equiv d/\Phi \gtrsim 0.1$ m the mean free path, and $\Phi$ the volume fraction.

For various $B$ at fixed $N$, the probability density functions (PDF) of $A$ show exponential tails that collapse on a single curve when rescaled by $\overline{A}$ (see top inset of Fig.\ \ref{fig06}). Similar results are found at fixed $B$ regardless of $N$. Moreover, since $A=v/\delta t$, the tail of the velocity's PDF is thus found to scale as $\exp(- c\  v / \overline{v} )$ independently of the volume fraction since $\overline{v} \sim h^0N^0B^{1/2}$. Consistently, the time lag distribution is found to scale as $\exp(- c' \tau / \overline{\tau})$, $c$ and $c'$ being dimensionless constants.

\section{Discussion} We have obtained the equation of state of a dissipative granular gas driven stochastically by injecting rotational energy into each particle. With usual notations (the pressure $P$ on the lid $\sim Mg/S$, and the container volume $V=Sh$), the equation of state of Eq.\ (\ref{EOStheo}) thus reads

\begin{equation}
PV \sim NE_c\frac{L}{h}\ {\rm ,}
\end{equation}
with $E_c \sim \langle v^2 \rangle \sim B$ the mean translational kinetic energy per particle of velocity $v$. Surprisingly, this equation is close of the equation of state of a perfect gas ($PV=NE_c$) with a geometric correction: the particle-container length ratio. This can be partially ascribed to particle-wall interactions since the Knudsen number $Kn \gtrsim 1$. Moreover, it differs from the equation of state of a dissipative granular gas driven by a vibrating wall $PV\sim E_c$ with $E_c \sim \mathcal{V}^{\theta(N)}$ with $\mathcal{V}$ the forcing velocity of the wall, and $\theta(N)$ a decreasing function from $\theta=2$ at low $N$ to $\theta \simeq 0$ at large $N$ when the clustering phenomenon occurs \cite{Falcon99b}. Here, no clustering is observed even when the volume fraction is increased up to 40\%. To our knowledge, no clustering instability has been also observed in numerical simulations of dissipative granular gases that are only driven by rotational degrees of freedom \cite{Cafiero02}. 
%The fundamental reason is that magnetic energy is {\it continuously} injected in each particle and is balanced in average by the dissipated energy during collisions. For the boundary-forcing case, particles far from the vibrating wall are colder than the other, and if their local density increases, it leads to an increase of the local collision frequency, and thus of the local dissipation, which in turn increases their local density.}
 We also show that the magnetic field $B$ in our experiment is the analogous of the thermodynamic temperature for molecular gases, or the analogous of the granular temperature for dissipative granular gases since $\langle v^2 \rangle \sim B$. The distribution of particle velocity near the top wall displays an exponential tail and is independent of the particle density. It is thus not Gaussian as for an ideal gas, or stretched exponential and density dependent as for a boundary-forced granular gas \cite{Rouyer00}.  Finally,  the collision frequency $\sim 1/ \langle \tau \rangle$ is found to scale as $N\sqrt{B}$. This result is consistent with the collision frequency of ideal gases $\sim N \sqrt{\langle v^2 \rangle}$, but not with the one of vibro-fluidized dissipative granular gases in dilute regime $\sim N^{1/2}\mathcal{V}$ \cite{Falcon06}. This difference is related to the spatially homogeneous nature of forcing. 

\section{Conclusion}
We have experimentally studied for the first time a 3D granular gas driven stochastically by injecting rotational energy into each particle. This differs from previous experimental studies of granular gas where the energy was injected by vibrations at a boundary. The equation of state is experimentally identified and the collision statistics measured (distribution of velocity, scalings of the particle rms velocity and mean collision frequency with the forcing). Several differences are reported with respect to thermodynamiclike gas and/or non equilibrium vibro-fluidized dissipative granular gas: (i) the gas-like state equation has a geometric correction (container-particle length ratio), (ii) no cluster formation occurs at high density, and (iii) the particle velocity distribution displays an exponential tail. The use of this new type of forcing will be of primary interest to test experimentally the hypothetical equipartition of rotational and translational energy, a feature not guaranteed for out-of-equilibrium systems \cite{Nichol12}.    

\acknowledgments
We thank C. Wilhelm and F. Gazeau for fruitful discussions. This work has been supported by ESA Topical Team on granular materials N$^{\circ}$4000103461.

%%%%%%%%%%%%%%%%%%%%%%%%%%%%%%%%%%%%%%
%%%%%%%%%%%% REFERENCES %%%%%%%%%%%%%%%%%%
%%%%%%%%%%%%%%%%%%%%%%%%%%%%%%%%%%%%%%

\end{document}